\documentclass[twocolumn,aps,prl,showpacs,amssymb,floatfix]{revtex4}
\usepackage{graphicx}
\usepackage{bm}
\begin{document}
\title{Lagrangian statistics in fully developed turbulence}
\author{L.~Biferale$^1$, G.~Boffetta$^2$, A.~Celani$^3$,
A.~Lanotte$^4$, and F.~Toschi$^5$} \affiliation{ $^1$ Dipartimento di
Fisica and INFM, Universit\`a degli Studi di Roma ``Tor Vergata'', Via
della Ricerca Scientifica 1, 00133 Roma, Italy \\ $^2$ Dipartimento di
Fisica Generale and INFM, Universit\`a degli Studi di Torino, Via
Pietro Giuria 1, 10125, Torino, Italy \\ $^3$ CNRS, INLN, 1361 Route
des Lucioles, 06560 Valbonne, France \\ $^4$ CNR-ISAC,
Str. Prov. Lecce-Monteroni km.1200, 73100 Lecce, Italy\\ $^5$ Istituto
per le Applicazioni del Calcolo, CNR, Viale del Policlinico 137, 00161
Roma, Italy} \date{\today}
\begin{abstract}
The statistical properties of fluid particles transported by a fully
 developed turbulent flow are investigated by means of high resolution
 direct numerical simulations. Single trajectory statistics are
 investigated in a time range spanning more than three decades, from
 less than a tenth of the Kolmogorov timescale up to one large-eddy
 turnover time. Acceleration and velocity statistics show a neat
 quantitative agreement with recent experimental results. Trapping
 effects in vortex filaments give rise to enhanced small-scale
 intermittency on Lagrangian observables.
\end{abstract}

\maketitle The knowledge of the statistical properties of particle
tracers advected by a fully developed turbulent flow is a key
ingredient for the development of stochastic Lagrangian models in
such diverse contexts as turbulent combustion, pollutant dispersion, cloud
formation and industrial mixing \cite{P94,Pope,S01,Y02}.  Given the
importance of this problem, there are comparatively few experimental
studies of turbulent Lagrangian dispersion. Progress in this direction
has been hindered mainly by the presence of a wide range of dynamical
timescales, an inherent property of fully developed turbulence.
Indeed, in order to obtain an accurate description of the particle
statistics it is necessary to follow their paths with very high
resolution, i.e., well below the Kolmogorov timescale $\tau_{\eta}$,
and for a long time lapse -- of the order of an eddy turnover time
$T_L$.  The ratio of these timescales can be estimated as
$T_L/\tau_{\eta} \sim R_\lambda$ where the microscale Reynolds number
$R_\lambda$ ranges in the hundreds for typical laboratory
experiments. A recent breakthrough has been made by La Porta {\em et
al} \cite{LVCAB01,VLCBA02} who, borrowing techniques from high-energy
physics, were able to track three-dimensional
trajectories with a resolution of $0.05 \tau_\eta$ and thus to study
the statistics of particle acceleration. However, owing to the small
measurement volume, trajectories could be followed only up to a few
$\tau_\eta$, preventing any investigation of the long time
correlations along particle paths. Conversely, the acoustical
technique adopted by Mordant {\em et al} \cite{MMMP01} enabled them to
successfully track particles for durations comparable to $T_L$ but could
not  access time delays of the order of $\tau_\eta$, and was
restricted to one-dimensional measurements. More conventional
techniques, e.g. based on CCD cameras \cite{OM00}, reveal useful only
for moderate Reynolds numbers ($R_{\lambda} \simeq 100$) due to their
limitations in the acquisition rate. In addition, all laboratory
techniques require a considerable amount of signal processing to get
rid of experimental noise. Besides these drawbacks, increasing
difficulties are met when attempting to deal with multi-particle
tracking.

In the light of the above remarks Direct Numerical Simulation of fully
developed turbulence represents a valuable alternative tool for the
investigation of Lagrangian statistics
\cite{RP74,YP89,SE91,Y91,VY99,BS02,IK02,CRLMPA03,SYBVLCB03} and its
important contribution has been recently reviewed in Ref.~\cite{Y02}.
Nonetheless, computational approaches have to face three demanding
requirements: (i) have the largest possible value of $R_\lambda$ in
order to maintain the flow in a fully developed turbulent state; (ii)
resolve properly the dissipative scales with the aim of computing
accurately small-scale observables; (iii) follow particle trajectories
for times comparable to a large-eddy-turnover time.  High resolution
and huge computational resources are therefore needed to accomplish
such a goal, making massive parallel computing by far and away the
most appropriate tool. In this Letter we report the results of Direct
Numerical Simulations of Lagrangian transport in homogeneous and
isotropic turbulence on the parallel computer IBM-SP4 at CINECA, on
$512^3$ and $1024^3$ cubic lattices with Reynolds numbers up to
$R_\lambda \sim 280$, a very accurate resolution of dissipative
scales, and a long integration time $T \approx T_L$.  The
Navier-Stokes equations
\begin{equation}
\partial_t {\bm u} + {\bm u} \cdot {\bm \nabla} {\bm u} = - {\bm
  \nabla} p + \nu \Delta {\bm u}\;,
\label{eq:1}  
\end{equation}
are integrated on a triply periodic box by means of a fully dealiased
pseudospectral code (the numerical parameters are listed in
Table~\ref{tab1}).
\begin{table*}
\begin{tabular}{cccccccccccc}
$R_\lambda$ & $u_{rms}$ & $\varepsilon$ & $\nu$ & $\eta$ & $L$ & $T_L$
& $\tau_\eta$ & $T$ & $\delta x$ & $N^3$ & $N_p$ \\ \hline 183 & 1.5 &
0.886 & 0.00205 &0.01 & 3.14 & 2.1 & 0.048 & 5 & 0.012 & 512$^3$ &
0.96$\cdot 10^{6}$ \\ 284 & 1.7 & 0.81 & 0.00088 &0.005& 3.14 & 1.8 &
0.033 & 4.4 & 0.006 & 1024$^3$ & 1.92 $\cdot 10^{6}$ \\ \hline
\end{tabular}
\caption{Parameters of the numerical simulations. Microscale Reynolds
number $R_\lambda$, root-mean-square velocity $u_{rms}$, energy
dissipation $\varepsilon$, viscosity $\nu$, Kolmogorov lengthscale
$\eta=(\nu^3/\varepsilon)^{1/4}$, integral scale $L$, large-eddy
turnover time $T_L = L/u_{rms}$, Kolmogorov timescale
$\tau_\eta=(\nu/\varepsilon)^{1/2}$, total integration time $T$, grid
spacing $\delta x$, resolution $N^3$, and the number of Lagrangian
tracers $N_p$.}
\label{tab1}
\end{table*}
Energy is injected at an average rate $\epsilon$ by keeping constant
the total energy in each of the first two wavenumber shells
\cite{CDKS93}. With the present choice of parameters the dissipative
range of lengthscales is extremely well resolved. With regard to the
scaling properties of the velocity field, our results are in perfect
numerical agreement with previous numerical simulations at comparable
$R_\lambda$ (see e.g. \cite{KIYIU03}). Upon having reached
statistically stationary conditions for the velocity field, millions
of Lagrangian tracers have been seeded into the flow and their
trajectories integrated according
\begin{widetext}
\begin{figure*}[ht]
\includegraphics[draft=false,scale=0.25]{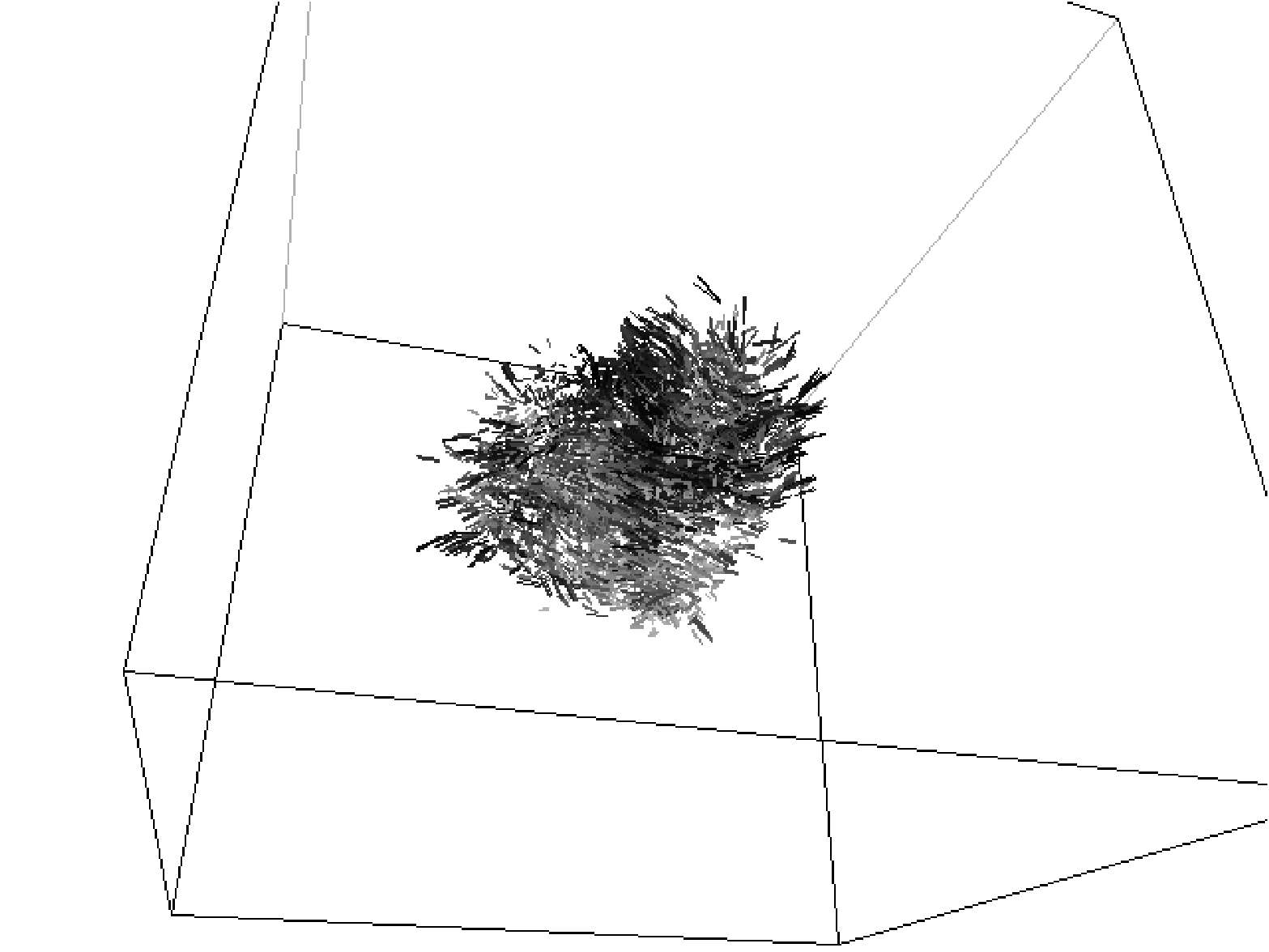}
\includegraphics[draft=false,scale=0.25]{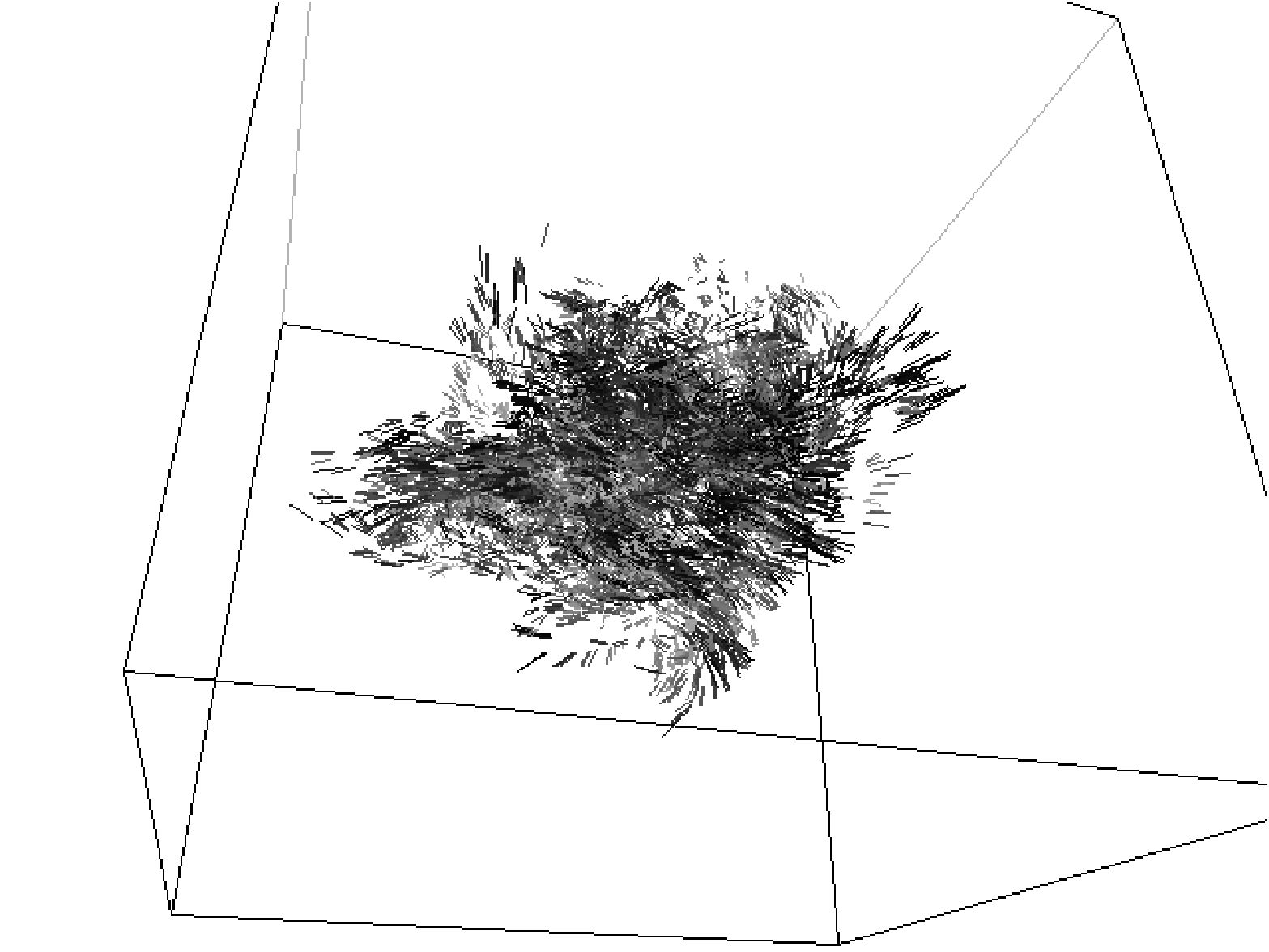}
\includegraphics[draft=false,scale=0.25]{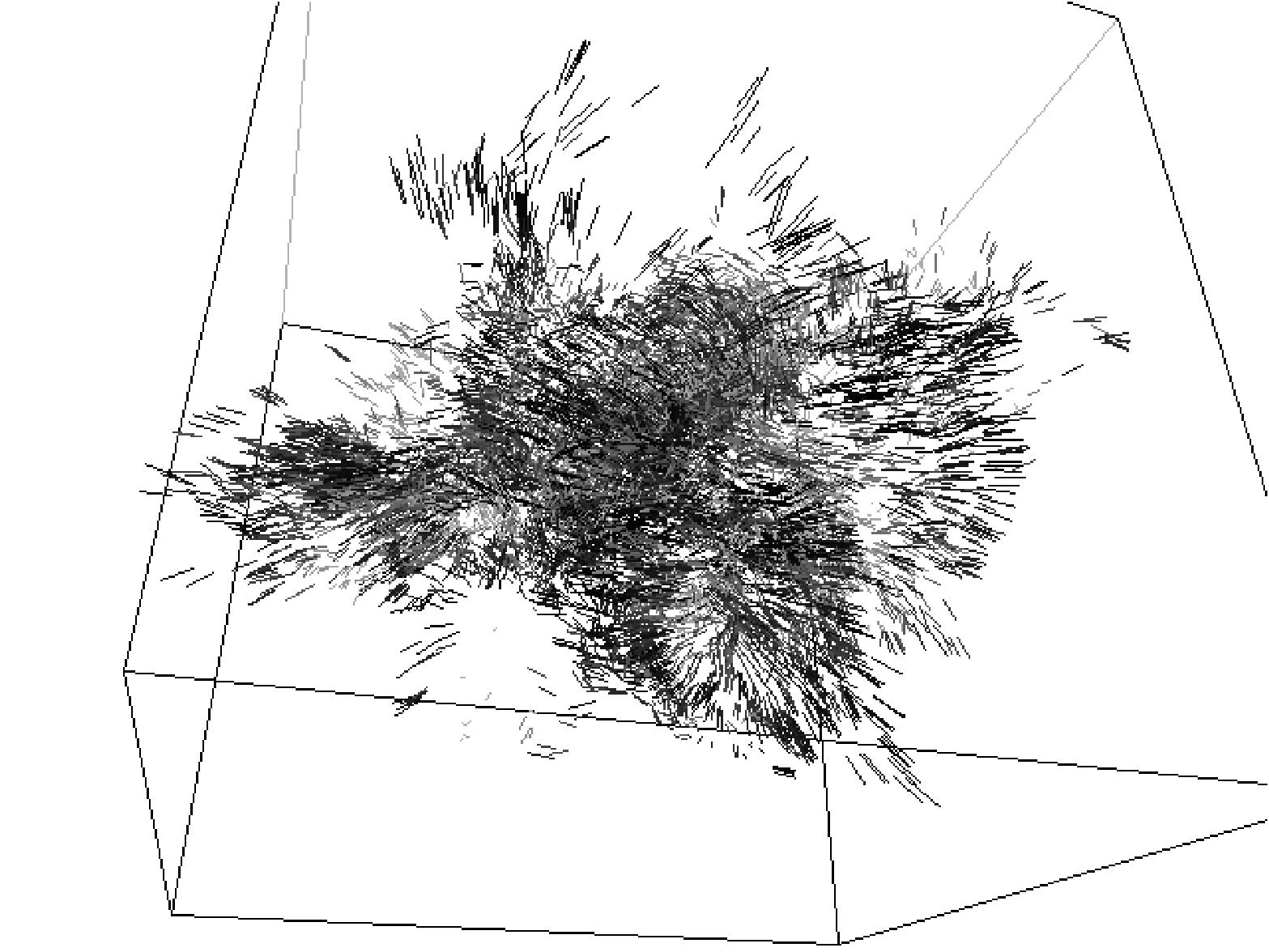}
\caption{Snapshots of a subset of Lagrangian particles ( $\approx
12000$) at three different times $t=(0.2, 0.5, 0.8)T_L$.  Gray
scales encode different velocity amplitudes.  }
\label{fig:1}
\end{figure*}
\end{widetext}
to
\begin{equation}
\frac{d {\bm X}}{dt} = {\bm u}({\bm X}(t),t)\;,
\end{equation}
over a time lapse of the order of $T_L$ (See Fig.~\ref{fig:1}).
Particle positions ${\bm X}(t)$ and velocities ${\bm v}(t)=\dot{\bm
X}(t)$ have been stored at a sampling rate $0.07 \tau_\eta$. The
forces acting on the particle -- pressure gradients ${\bm \nabla}p
({\bm X}(t),t)$ , viscous forces $\nu\Delta {\bm u} ({\bm X}(t),t)$
and external input -- and the resulting particle acceleration ${\bm
a}(t)=\dot{\bm v}(t)$ have been recorded along the particle paths
every $0.14 \tau_\eta$. The resulting database permits detailed
study of the statistics of Lagrangian velocity and acceleration
over a range of timescales spanning more than three decades.

\begin{figure}[hbt]
\includegraphics[draft=false,scale=0.68]{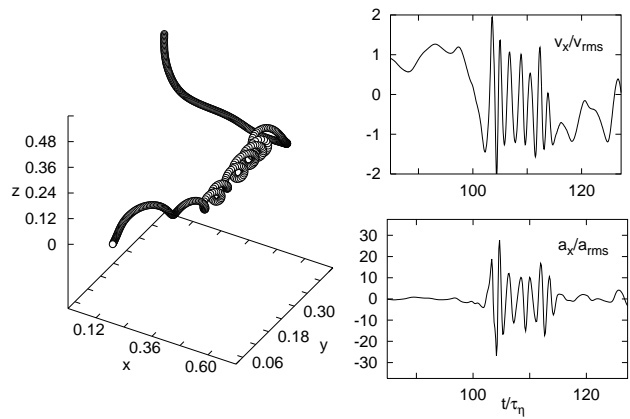}
\caption{Trajectory and time series. Left panel: 3D trajectory of a
trapping event in vortex filament. Acceleration and velocity
fluctuations here reach about $30$ and $2$ r.m.s. values, respectively
(right panels).}
\label{fig:2}
\end{figure}

In what follows,  we will focus on the description of the statistics of
single particle trajectories, deferring the discussion of
multi-particle statistics to a forthcoming publication.  The analysis
reveals the presence of frequent entrapment events within vortical
structures (see Fig.~\ref{fig:2}). The characteristic frequency of
these spiraling paths is comparable to $\tau_\eta^{-1}$ and their
duration can be as long as $10 \tau_\eta$. The velocity experienced
during these events can attain $5 u_{rms}$ with an ensuing
acceleration $5 u_{rms} / \tau_\eta$ as large as $80 a_{rms}$.  As we
will show below, these events have a significant impact on time
correlations up to $10 \tau_\eta$.
\begin{figure}[hbt]
\includegraphics[draft=false,scale=0.68]{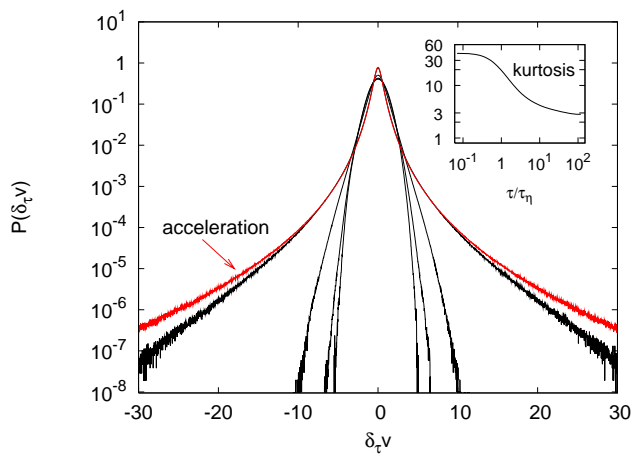}
\caption{Probability density functions of velocity increments and
acceleration, normalized with their variance. Curves refer to time
increments $\tau = (97, 25, 6, 0.7)\tau_{\eta}$ from inside to
outside, and to the acceleration (outermost curve). In the inset, the kurtosis
$K(\tau)=\left\langle{\delta_{\tau} v}^4\right\rangle/(\left\langle
\delta_{\tau} v)^2\right\rangle^2 $ for the entire time interval
$0.07\tau_{\eta} \div 2 T_L$. The saturation of $K(\tau)$ at small
time increments is an indication of the high numerical resolution.}
\label{fig:3}
\end{figure}
In Fig.~\ref{fig:3} we present the probability density functions
(PDFs) of velocity differences $\delta_\tau v = v(t+\tau)-v(t)$ taken
along a given direction and rescaled by their variance $\langle
(\delta_\tau v)^2\rangle^{1/2}$. For the shortest time delays shown in
the figure the velocity increment PDF approaches the distribution of
the acceleration $a(t)$ which is characterized by a strong
intermittency, stretched exponential tails and a flatness $\langle
a^4\rangle/ \langle a^2 \rangle^2 \approx 40$. Classical Kolmogorov
scaling arguments (see e.g. Ref.~\cite{MY75}) yield for the
acceleration variance the prediction
\begin{equation}
  \langle a^2 \rangle = a_0 \,\varepsilon^{3/2} \nu^{-1/2}\;.
\end{equation}
We measure the constant $a_0=3.5 \pm 0.3$, in agreement with the
results of Ref.~\cite{VY99} at comparable $R_\lambda$. Yet, one needs
to notice that intermittent corrections may introduce in the
definition of $a_0$ some dependency on Reynolds number\cite{VLCBA02}.  At
larger time separations the PDFs are decreasingly intermittent and
eventually become slightly sub-Gaussian for $\tau \approx T_L$, with a
flatness $\langle(\delta_\tau v)^4\rangle/ \langle (\delta_\tau v)^2
\rangle^2 \approx 2.8$. Assuming a scaling law for the second-order
Lagrangian velocity structure function in the time-range $\tau_\eta
\ll \tau \ll T_L$, dimensional analysis \cite{MY75}, predicts $\langle
(\delta_\tau v)^2 \rangle=C_0 \varepsilon \tau$.  Remark that this
prediction is not affected by intermittency corrections owing to its
linear dependence on the energy dissipation rate \cite{N89,B93,BDM02}.
Therefore, $C_0$ is expected to be universal.  For our data the
scaling behavior holds on a relatively narrow window in $\tau$,
leading to some uncertainty in the determination of the numerical
prefactor. We estimate $C_0 = 5 \div 6$, not far from previous
measurements \cite{MMMP01,LD02}.

\begin{figure}[htb]
\centering
\includegraphics[draft=false, scale=0.68]{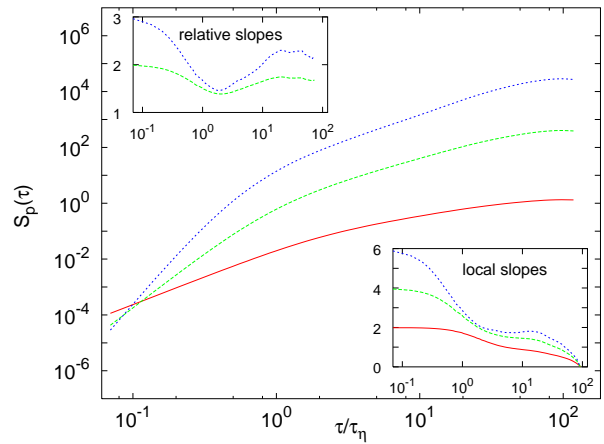}
\caption{Log-log plot of Lagrangian structure functions of orders
 $p=2,4,6$ (bottom to top) {\em vs} $\tau$. Bottom right: logarithmic
 local slopes $d \log S_p(\tau)/ d \log \tau$ (same line styles).  Top
 left: relative local slopes with respect to the second order
 structure function $d \log S_p(\tau) / d \log S_2(\tau)$, for
 $p=4,6$. Data refer to the $v_x$ component. The two other
 velocity components exhibit slightly worse scaling due to anisotropy
 effects.}
\label{fig:4}
\end{figure}

The intermittency in the PDFs of $\delta_\tau v$ can be conveniently
quantified in terms of the Lagrangian velocity structure functions
$S_p(\tau)= \langle (\delta_\tau v)^p \rangle$, that are expected to
behave as power laws $\tau^{\xi_p}$.  The scaling exponents $\xi_p$
can be evaluated by looking at the logarithmic slope $d \log S_p(\tau)
/ d \log \tau$ that should display a plateau in the range $\tau_\eta
\ll \tau \ll T_L$. From Fig.~\ref{fig:4} it is clear that it is very
difficult to extract the values for $\xi_p$. However, by means of the
extended self-similarity procedure \cite{BCTBMS93}, it is possible to
estimate the relative exponents $\xi_4/\xi_2=1.7\pm0.05$,
$\xi_5/\xi_2=2.0\pm0.05$, $\xi_6/\xi_2=2.2 \pm0.07$, in fair agreement
with those obtained in Ref.~\cite{MMMP01}. The range of time delays
over which relative scaling occurs is $10\tau_\eta$ to
$70\tau_\eta$. In this range anisotropic contributions induced by the
large scale flow appear to influence the scaling properties.

\begin{figure}[htb]
\centering
\includegraphics[draft=false, scale=0.68]{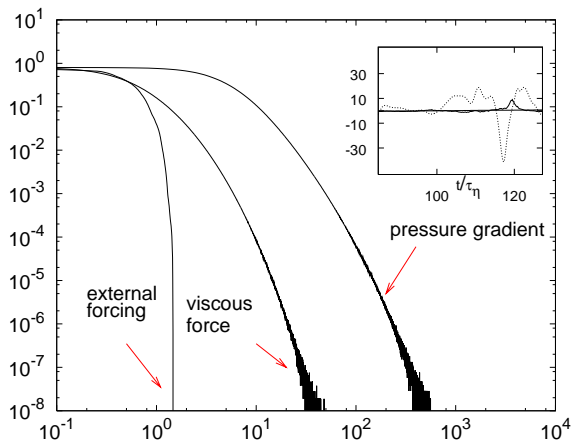}
\caption{Log-log plot of PDFs for $-\partial_x p$, $\nu \Delta u_x$,
$F_x$.  The external forcing is virtually negligible, and the main
contribution to large accelerations is made by pressure gradients.
Inset: a typical evolution of the three terms along a particular
trajectory. The strongest signal is $\partial_x p$ (dashed line),
while the viscous force is activated only as a subleading response to
pressure gradients (solid line). The force contribution is
indistinguishable from zero.}
\label{fig:5}
\end{figure}

It is interesting to remark that for values
of $\tau$ in the range from $\tau_\eta$ to $10 \tau_\eta$, the local slopes are
significantly smaller and tend to accumulate around the value $2$ for
all orders. This relevant correction to scaling cannot be attributed
to the influence of the dissipative range $\tau < \tau_\eta$, since
the latter would increase the value of the local slope, rather than
decreasing it. A similar effect can be detected in Eulerian structure
functions as well, yet the intensity in the latter case is much less
pronounced.  These strong deviations in the Lagrangian scaling laws
are most likely due to the trapping events depicted in
Fig.~\ref{fig:2}. Indeed, the long residence time within small-scale
vortical structures introduces an additional weighting factor that
enhances the effect with respect to Eulerian measurements.
Further insight into the single particle dynamics can be gained by
investigating the various force terms concurring to determine the
total particle acceleration
$$
 \frac{d^2{\bm X}(t)}{dt^2}=-{\bm \nabla}p({\bm X},t)+ \nu \Delta{\bm
u}({\bm X}(t),t)+ \bm F({\bm X}(t),t)\;.
$$
As shown in the inset of Fig.~\ref{fig:5}, it is evident that the main
source of acceleration is the pressure gradient \cite{VY99}, whereas
the viscous term is important only near relatively strong  acceleration events
and the forcing  contribution is negligible. A more quantitative test is
given by the comparison of the PDFs of the three forces in Fig.~\ref{fig:5}. 
To conclude, we have presented the analysis of single-particle
statistics in high Reynolds number flows. Our results fit well with
previous experimental measurements. At variance with experiments, we
can investigate the statistical properties of millions of particles
on a wide range of time intervals,  from a small fraction of the
Kolmogorov time up to the integral correlation time. We found  clear
indications that velocity fluctuations along Lagrangian trajectories
are affected by multiple-time dynamics. In the interval $10\tau_{\eta}
<\tau < T_L$ we observed anomalous scaling for Lagrangian velocity
structure functions. For frequencies of the order of
$\tau_{\eta}^{-1}$ we noticed that velocity fluctuations are affected
by events where particles are trapped in vortex filaments. Events
with trapping times much longer than expected on the basis of simple
dimensional analysis appear frequently. The main novelty of Lagrangian
single-particle velocity fluctuations with respect to Eulerian
fluctuations is  probably  the capture of trajectories induced by
small-scale structures. For example, the event analyzed in
Fig.~\ref{fig:2} would have a much smaller weight in an Eulerian
analysis because of large-scale sweeping past the fixed probe.
Among the most challenging open
problems arising from our analysis are  how to
incorporate such dynamical processes in stochastic modelization of
particle diffusion \cite{S01} and in the Lagrangian multifractal
description \cite{B93,BDM02}.
\begin{acknowledgments}
The simulations were performed within the keyproject ``Lagrangian Turbulence''
on the IBM-SP4 of  Cineca (Bologna,
Italy). We are grateful to C.~Cavazzoni and G.~Erbacci for resource
allocation and precious technical assistance.  We acknowledge support
from EU under the contracts HPRN-CT-2002-00300 and HPRN-CT-2000-0162.
We also thank E. L\'ev\^eque for useful discussions and
 B. Devenish for a careful reading of the manuscript.
\end{acknowledgments}                              
                             

\end{document}